\documentclass[12pt,a4paper]{article}
\usepackage[utf8]{inputenc}
\usepackage[T1]{fontenc}
\usepackage{fix-cm}
\usepackage{amsmath,amssymb}
\usepackage{graphicx}
\usepackage{dcolumn}
\usepackage{bm}
\usepackage[
    colorlinks=true,
    linkcolor=blue,
    citecolor=blue,
    urlcolor=blue,
    filecolor=blue
]{hyperref}
\usepackage{doi}
\usepackage{geometry}
\usepackage{authblk}
\usepackage[numbers,sort&compress]{natbib}
\providecommand{\keywords}[1]{\noindent\textbf{Keywords:} #1}
\providecommand{\pacs}[1]{\noindent\textbf{PACS:} #1}

\newenvironment{acknowledgments}{\section*{Acknowledgments}}{}

\begin{document}

\title{Electrons with Anomalous Energy Generated in Vacuum Diodes with Different Pulse Duration}
\author[1]{Vasily~Yu.~Kozhevnikov\thanks{\texttt{Vasily.Y.Kozhevnikov@ieee.org}}}
\author[2]{Victor~F.~Tarasenko\thanks{\texttt{vf.tarasenko@hcei.ru}}}
\author[3]{Maxim~S.~Vorobyov}
\author[2]{Evgenii~Kh.~Baksht}
\author[1]{Andrey~V.~Kozyrev}

\affil[1]{Laboratory of Theoretical Physics, Institute of High Current Electronics, Tomsk 634055, Russian Federation}
\affil[2]{Laboratory of Optical Radiations, Institute of High Current Electronics, Tomsk 634055, Russian Federation}
\affil[3]{Laboratory of Plasma Emission Electronics, Institute of High Current Electronics, Tomsk 634055, Russian Federation}

\date{\today}

\maketitle

\begin{abstract}
This paper presents results from an experimental study in which anomalous energy electrons (AEE) appear in vacuum diodes under varying amplitude and pulse-duration conditions. AEE refer to high-energy discharge electrons with kinetic energy $T$ exceeding $eU$ values (where $e$ is the electron charge and $U$ is the amplitude gap voltage). Here, two experimental setups were used to generate electron beams with different pulse durations. The electron energy spectra were reconstructed from the beam attenuation curves using an AI-driven methodology for solving an ill-posed inverse problem for the Fredholm equation. This confirms that vacuum diodes generate electron beams with a significant AEE fraction (up to 25~\%) only when operating with nanosecond-long voltage pulses. It has been confirmed that AEE in vacuum is generated using nanosecond- or sub-nanosecond voltage pulses. Experiments also indicate that photoelectric absorption of bremsstrahlung cannot produce significant AEE in vacuum discharges.
\end{abstract}

\maketitle

\section{\label{sec:level1}Introduction}

Vacuum-diode accelerators with large-area cathodes (hundreds of cm\textsuperscript{2}) are widely used in laser development \cite{Basov_1993, Sethian_1997, Bychkov_2000} and in technological processes \cite{Livesay_1993, Kozawa_2004}. In these applications, beam-current pulse durations are of the order of microseconds or longer. In these discharges (e.g. \cite{Basov_1993, Sethian_1997, Bychkov_2000, Livesay_1993, Kozawa_2004}) the electrons with anomalously high energy (AEE) are not registered. Many studies have recorded these electrons (AEE) in vacuum \cite{Khudyakova, Bugaev, Shpak, Ganter_2008} and gas \cite{Tarasova, Babich, Baksht_2010, Kozyrev_2015} discharges.

In 2023, \cite{Pasko_2023} proposed a mechanism of photoelectric absorption of bremsstrahlung (MPAB) for AEE beams generated in atmospheric discharges. \cite{Pasko_2023} described this mechanism in detail in \cite{Pasko_2024} for a vacuum diode and compared it with experimental results from gas-filled diodes reported in \cite{Khudyakova, Tarasenko_2020}. In the MPAB analysis in \cite{Pasko_2024}, it was shown that some X-ray quanta generated during electron bombardment of a metal anode return to the cathode and knock out new electrons. Moreover, some of these electrons have an energy of $eU$. These electrons are then re-accelerated in the discharge gap, and their energy exceeds $T > eU$ by a factor of two or more over several cycles. However, \cite{Pasko_2024} did not consider cathode shape and area, which influence the direction of secondary-electron propagation under X-ray radiation. Furthermore, when analyzing the effect of cathode-material atomic number on runaway-electron beam current, the authors did not account for changes in diode voltage when using cathodes of different materials; see \cite{Zhang_2013}. Also, the authors made no comparison with theoretical studies \cite{Boichenko_2011, Kozhevnikov_2022}, which proposed another mechanism for AEE generation in vacuum diodes, or with experiments \cite{Babich, Kozyrev_2015}, which obtained electrons with energy $T>eU$ in gas-filled and vacuum diodes. In \cite{Pasko_2024}, quantitative evidence showed that X-ray formation in laboratory discharges determines runaway-electron generation from the cathode material, due to photoelectric absorption of bremsstrahlung generated by runaway electrons bombarding the anode. However, data on the generation of electron beams in vacuum and gas-filled diodes obtained in 2026 \cite{Tarasenko_2026_1, Kozhevnikov_2026} and earlier \cite{Babich, Kozyrev_2015, Mesyats_2020, Huang_2025} show that the physical mechanism proposed in \cite{Pasko_2023, Pasko_2024}, where comparison was made with results obtained for gas-filled diodes \cite{Khudyakova, Tarasenko_2020}, is not the dominant one at electrode voltages of hundreds of kilovolts.

The objective of the present study is to experimentally investigate AEE generation in vacuum diodes under varying pulse durations and voltage amplitudes. The corresponding beam energy spectra are measured indirectly by using a newly developed AI-driven methodology (see \cite{Kozhevnikov_2026_plasma}), which substantially improves the accuracy of electron energy spectrum reconstruction. This paper continues that research; preliminary results are presented in \cite{Tarasenko_2026_1, Kozhevnikov_2026}.

\section{\label{sec:level2}Experimental setups and measurement techniques}

\begin{figure}[ht]
\centering
\includegraphics[scale=.26]{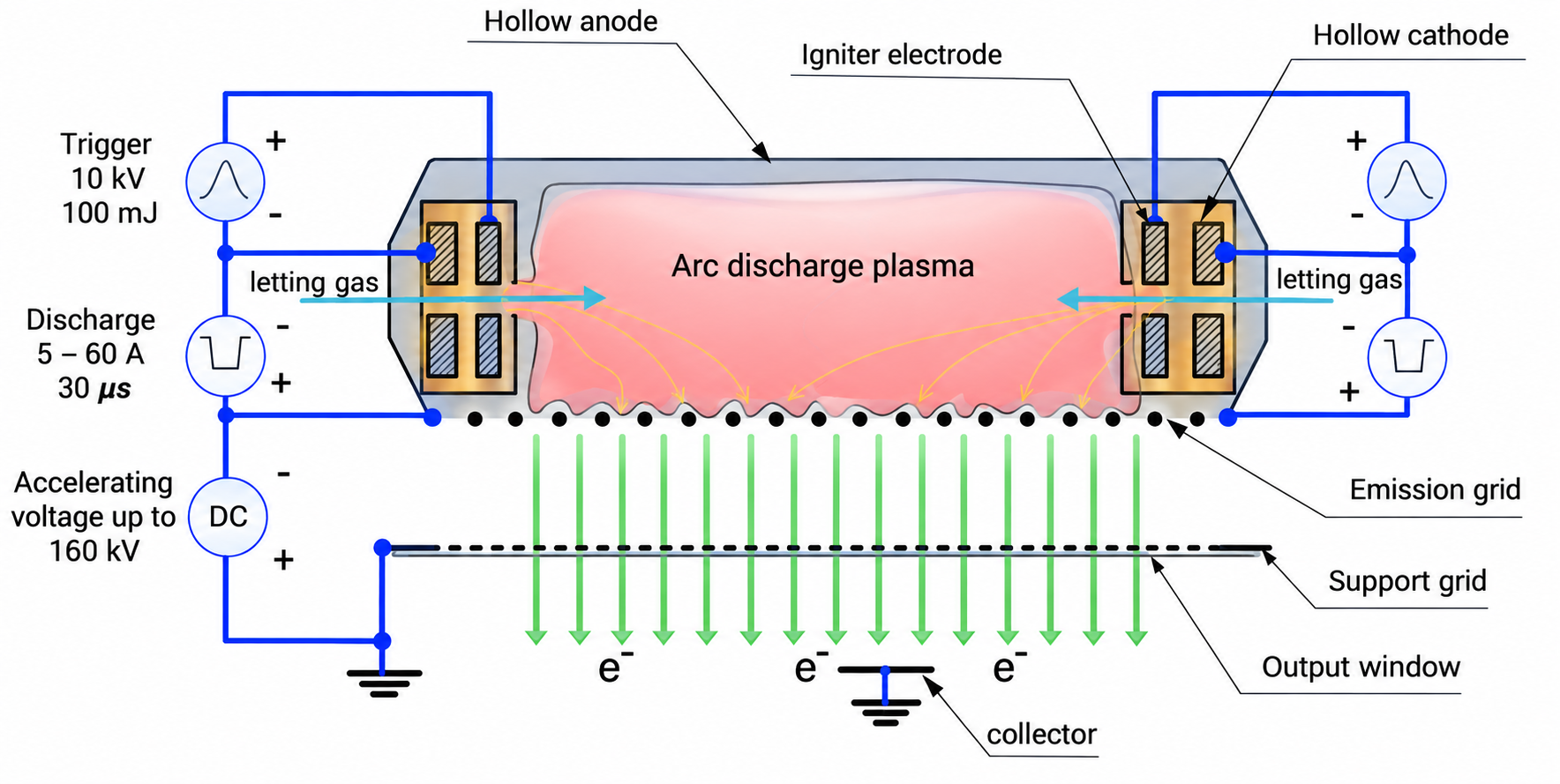}
\caption{\label{fig:1} Schematic of a constant-voltage accelerator showing the vacuum diode during beam current generation.}
\end{figure}

We investigated AEE beam generation using two experimental setups equipped with vacuum diodes. Setup~1 used the ``DUET'' pulsed electron accelerator, which features a plasma cathode and extracts a large-aperture electron beam ($75 \times 15$~cm) into atmospheric-pressure air. In this setup, the cathode and anode are flat, according to papers on MPAB \cite{Pasko_2023, Pasko_2024}. The accelerator's main characteristics and operating principle are described in \cite{Vorobyov_2015}. The corresponding setup schematic is shown in Fig.~\ref{fig:1}.

Under a constant accelerating voltage applied between the emission grid and the output foil window, which serves as the anode of the high-voltage diode gap, electrons are extracted from the plasma and accelerated to an energy corresponding to the accelerating voltage. The flat part of the cathode was covered with a grid having an optical transparency of 44~\% and measured $75 \times 15$~cm. Under the experimental conditions, the voltage droop across the gap did not exceed 3~\%. The electron beam is extracted into the atmosphere through an exit window measuring $75 \times 15$~cm, covered with a $40~\mu$m-thick foil made of AB-50 aluminum–beryllium composite. The foil was laid on a support structure with a geometric transparency of 56~\%. The gap voltage has been measured with a high-resistance voltage divider, and the total current through the diode has been recorded with a Rogowski coil installed on the positive pole of the capacitor bank. At 10~mm from the exit foil, a shielded collector with a 56~mm-diameter receiving part collects a portion of the electron current extracted from the diode gap and releases it into the atmosphere. The foil filter thickness was adjusted by stacking the required number of aluminum foils. As the foil-filter thickness increased, the collector current decreased, and this decrease continued until the main (useful) signal became comparable to the electrical interference level.

\begin{figure}[ht]
\centering
\includegraphics[scale=.175]{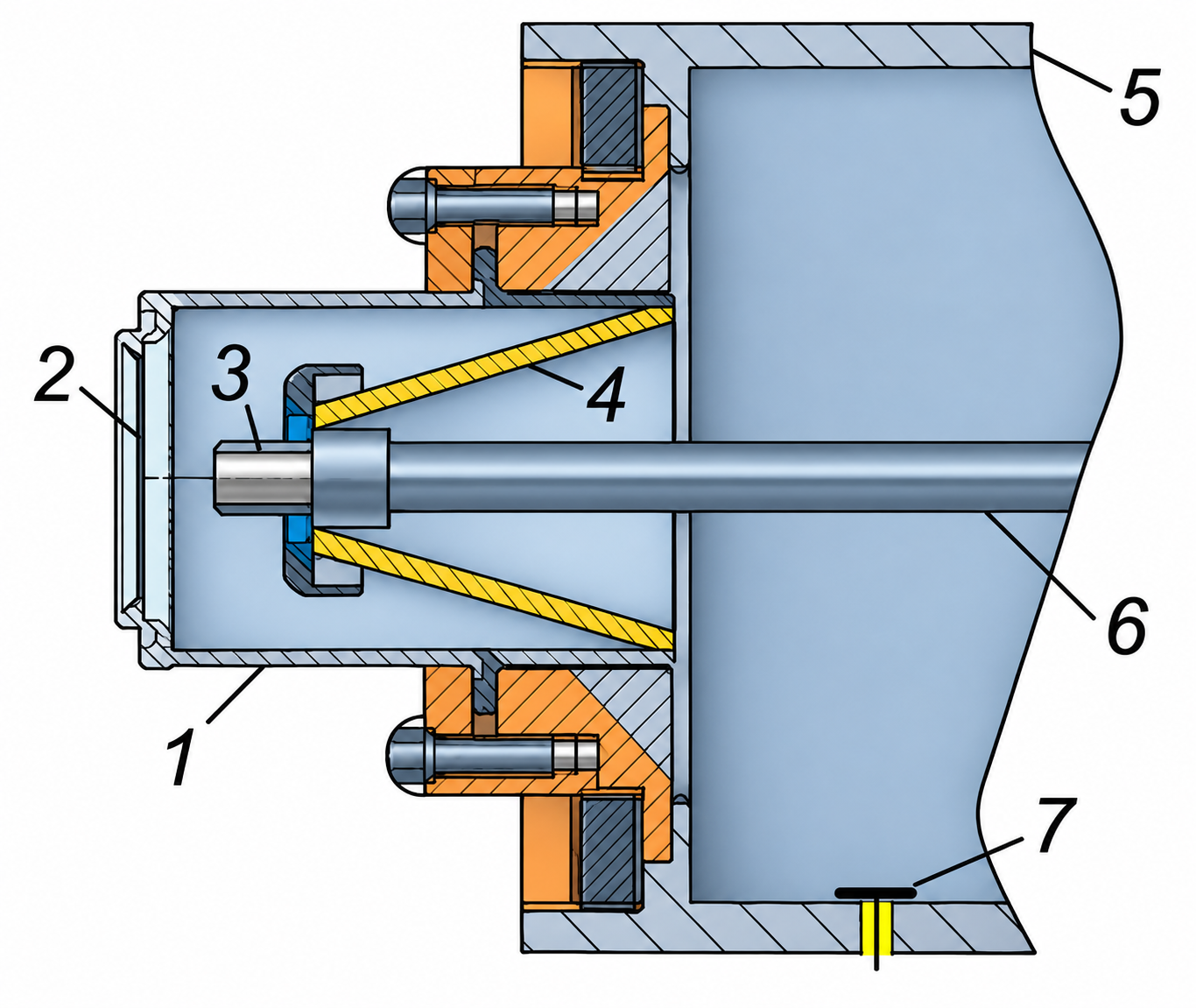}
\caption{\label{fig:2} The output section of the RADAN-220 generator with the IMA3-150E sealed-off vacuum tube at its left end: \textit{1} – electron tube housing; \textit{2} – beryllium foil window; \textit{3} – tubular cathode; \textit{4} – insulator; \textit{5} – part of the RADAN-220 generator housing; \textit{6} – high-voltage generator terminal immersed in transformer oil; \textit{7} – capacitive voltage probe.}
\end{figure}

As shown in \cite{Tarasenko_2026_1, Kozhevnikov_2026}, when using the SLEP-150M nanosecond generator with a vacuum diode and a cylindrical cathode 6~mm in diameter, the AEE fraction can exceed 15~\% of the electrons in the beam. To confirm the reliability of these results, measurements of the electron beam current were performed and its spectrum was calculated in Setup~2. Here, the RADAN\mbox{-}220 generator \cite{Yalandin_2001} produced negative-polarity voltage pulses with 2~ns FWHM, 0.5~ns rise time, and an amplitude of 200~kV. The experiments were carried out using a sealed-off IMA3-150E vacuum tube installed at the generator's end face (Fig.~\ref{fig:2}).

The electron-beam current was measured using a 20~mm-diameter collector connected to a cable with a characteristic impedance of 50~ohms. The collector’s temporal resolution was better than~80 ps. The voltage and beam-current pulses were fed to real-time digital oscilloscopes. To record current and voltage oscillograms on Setup~1, a TDS-3034 real-time oscilloscope (0.3~GHz and 2.5~samples/ns) was used. For operation on Setup~2, a Tektronix~TDS6604 oscilloscope (6~GHz and 20~samples/ns) was used. To record the beam-current pulses in this setup, a 1~m-long RG58-A/U (Radiolab) high-frequency cable and N-type (Suhner 11 N-50-3-28/133 NE) and SMA (Radiall R125.075.000) connectors were used. 

\section{\label{sec:level3}Results and discussion}
\subsection{Electron beam current measurements and spectrum reconstruction}

\begin{figure}[ht]
\centering
\includegraphics[scale=.2]{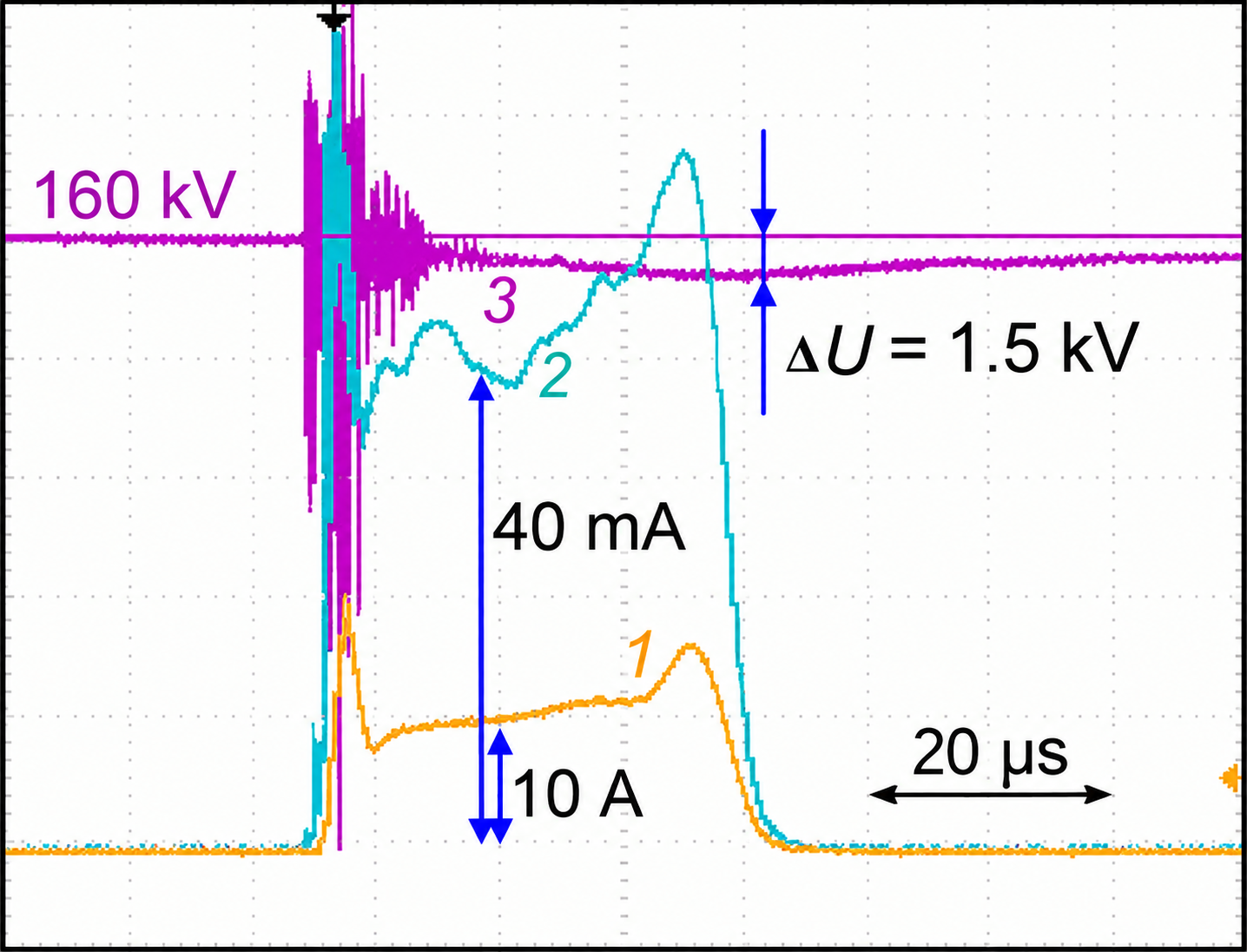}
\caption{\label{fig:3} Waveforms of the electron-beam current in the acceleration gap (\textit{1}), current at the collector after the beam passes through a foil filter of thickness 45~$\mu$m (\textit{2}), and voltage sag at the high-voltage capacitor bank (\textit{3}) at a vacuum diode voltage of 160~kV.}
\end{figure}

In \cite{Pasko_2024}, the diode geometry used in the MPAB analysis assumed two flat electrodes. Setup~1 (Fig.~\ref{fig:1}) most closely realizes these conditions. Here, the cathode and anode are flat and have large areas. A long beam current duration is also important for detecting MPAB \cite{Pasko_2024}. (Fig.~\ref{fig:3}) shows typical oscillograms of voltage pulses (\textit{1}), the collector current (\textit{2}) after the beam passes through a foil filter with a thickness of $45\ \mu$m, and the total current in the accelerating gap (\textit{3}).

\begin{figure}[ht]
\centering
\includegraphics[scale=.43]{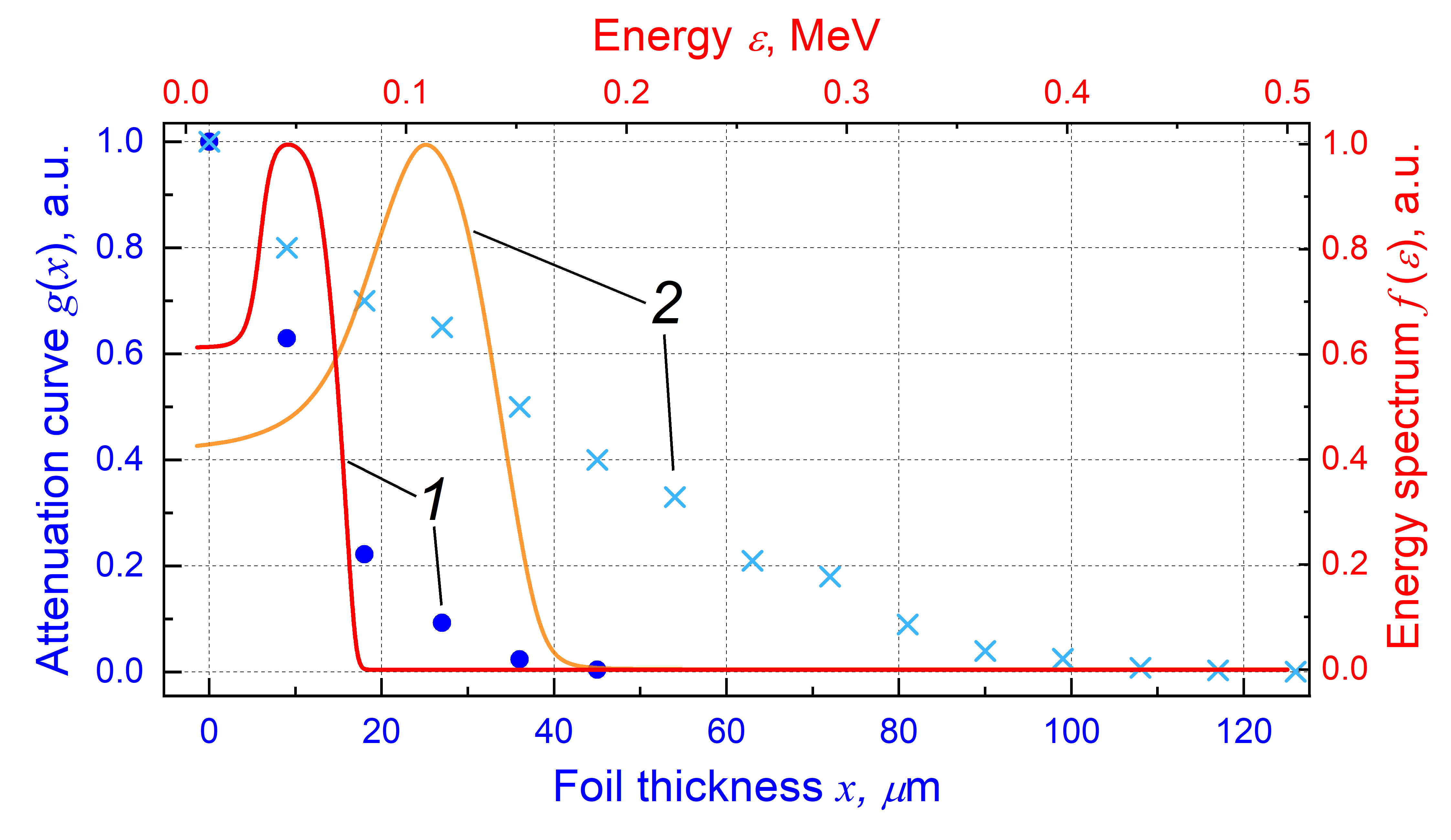}
\caption{\label{fig:4} Electron beam attenuation curves (dots and crosses) and electron energy spectra (solid lines) obtained with Setup~1 at anode voltages of 110~kV~(\textit{1}) and 160~kV~(\textit{2}).}
\end{figure}

Fig.~\ref{fig:3} shows the voltage oscillogram at an enlarged scale to highlight the small voltage drop across the high-voltage capacitor bank connected to the accelerating gap. The oscillations in the voltage oscillogram, whose amplitude did not exceed 23~kV, correspond to the electromagnetic pickup signal during breakdown between the igniter electrode and the cathode (Fig.~\ref{fig:1}). Fig.~\ref {fig:4} shows the attenuation curves for two operating voltages obtained under these conditions, along with the reconstructed electron spectra according to the previously proposed methodology \cite{Kozhevnikov_2026_plasma}.

In this study, two acceleration voltages differing by a factor of 1.5 were deliberately selected. The resulting electron-beam spectra indicate that, at acceleration potentials on the order of several hundred kilovolts, anomalous energy electrons (AEE) cannot be recorded reliably (with accuracy better than 1~\%). However, they show their trend changes and confirm the results of \cite{Tarasenko_2026_1, Kozhevnikov_2026}, respectively. With a relatively large effective cathode area and a vacuum-diode voltage of 160~kV, the AEE fraction does not exceed 0.7~\% and drops as the diode operating voltage decreases. 

\begin{figure}[ht]
\centering
\includegraphics[scale=.43]{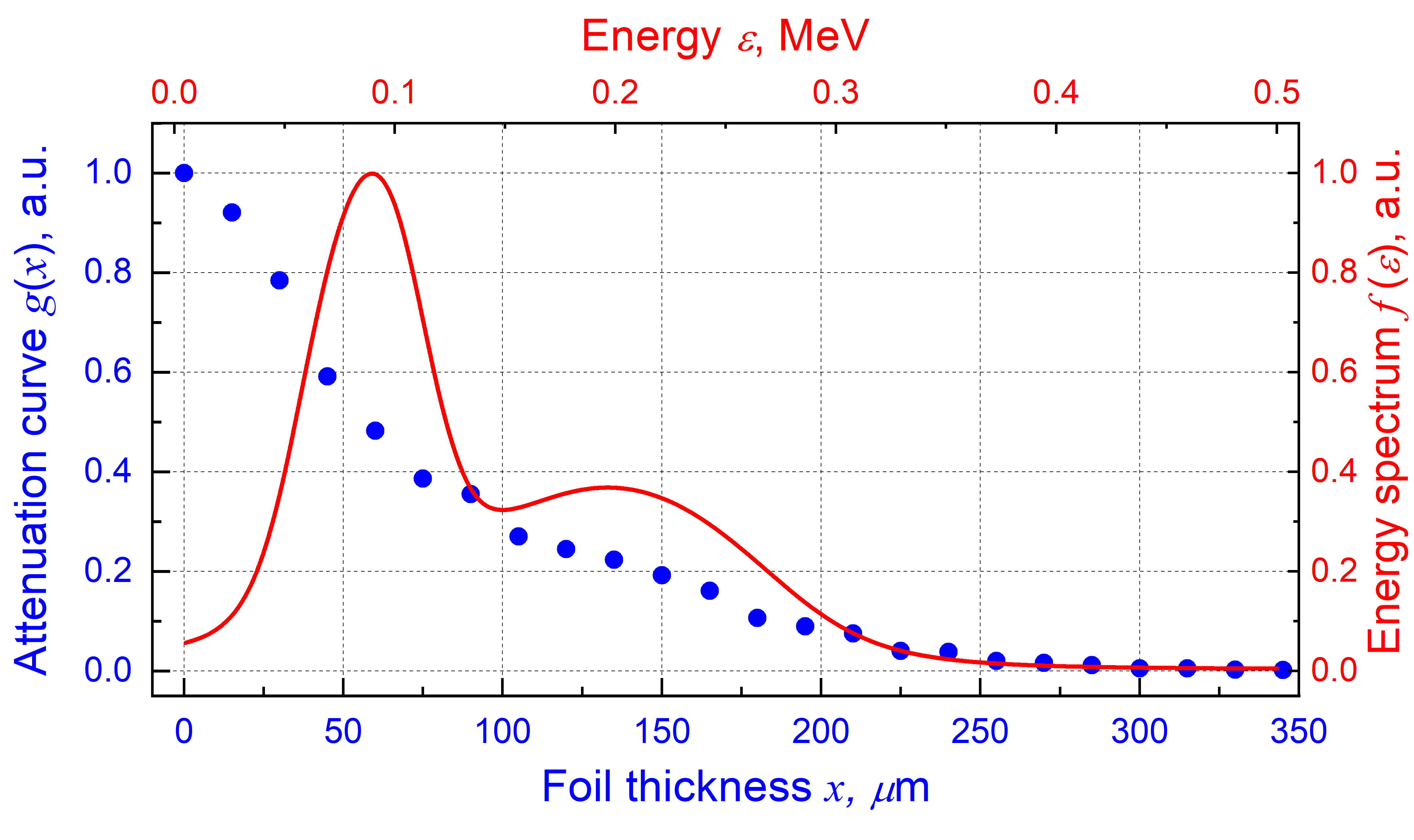}
\caption{\label{fig:5} Electron beam attenuation curve (dots) and electron energy spectrum (solid line) obtained at Setup~2 with an amplitude anode voltage of 200~kV.}
\end{figure}

Setup~2 used a generator with a nanosecond-duration voltage pulse. Fig.~\ref {fig:5} shows the obtained attenuation curve and the calculated electron energy spectrum. This spectrum shows two groups of electrons, with the second group having energy exceeding $eU$, accounting for 25~\%. As shown in \cite{Baksht_2007} and confirmed here, AEE generation occurs at the front edge of the beam current pulse.

\subsection{Discussion}

Understanding the AEE generation mechanism in vacuum and gas discharges of various types is important. AEEs affect most technological processes in industry and laboratories and require increased personnel protection during long-term accelerator operation. In addition, terrestrial gamma-ray bursts from thunderstorm activity on Earth \cite{Pasko_2023, Stephan_2021} are also associated with AEE generation.

In the theoretical paper \cite{Pasko_2024}, a physical mechanism was proposed in which the increase in electron energy is achieved through photoelectric absorption of bremsstrahlung generated during a previous cycle of electron bombardment of the anode. From the calculations performed in that work, it follows that at a gap voltage of $U = 100$~kV and
1~MV, the fraction of electrons with energy exceeding approximately their initial value 
$eU$ during the second pass is 0.85~\% and 8.48~\%, respectively. The total number of electrons with energy $T > eU$ during the third pass is 0.008~\% and 0.72~\%, respectively, and it continues to decrease rapidly thereafter. Although both cases form high-energy electrons exceeding $eU$ by about a factor of 2.5 over several passes, the average AEE value exceeds $eU$ by only 1-7~\%. Our data on electron-beam spectra obtained in Setup~1 with a long current pulse duration (tens of microseconds) do not contradict these results. With a relatively large effective cathode area of 500~cm\textsuperscript{2} and a vacuum diode voltage of 160~kV, the AEE fraction is small and does not exceed 0.7~\%. 

In contrast, the AEE fraction in the experiment on Setup~2 was 24~\%, which significantly exceeds the data from Setup~1 and those reported in \cite{Pasko_2024} for beam electrons with initial energies of 100~keV and 1~MeV. Furthermore, the AEE generation data from Setup~2 are consistent with theoretical calculations in \cite{Boichenko_2011, Kozhevnikov_2022}. Those works showed that in vacuum diodes, most AEE is generated at the front edge of the voltage pulse during rapid redistribution of the electric field and space charge in the diode. This occurs because of a significant mismatch between cathode emission inflow and electron outflow through the collector (anode) in high-current vacuum devices operating in the nanosecond- and sub-nanosecond-pulse range.

It should be noted that in \cite{Pasko_2024}, when comparing with experimental data on AEE generation in a gas diode, see the reference to \cite{Tarasenko_2020}, which is described in more detail in \cite{Zhang_2013}, the change in voltage across the gas diode with an increase in the atomic number of the cathode was not taken into account. In \cite{Zhang_2013}, the authors showed that the highest beam current amplitudes were obtained with a stainless steel cathode at a voltage pulse duration of 1~ns and a rise time of 0.3~ns. With a copper cathode, which has a higher atomic number than iron, cobalt, and nickel (the metals composing stainless steel), the beam current was 2–3.5 times lower. Increasing the voltage pulse amplitude with increasing atomic number was achieved by reducing electron emission from the cathode and increasing the voltage amplitude across the gas diode during runaway electron-beam generation. Accordingly, increasing the beam current amplitude with increasing atomic number of the cathode material was achieved by reducing the emissivity of the stainless steel cathode and, consequently, by increasing the voltage pulse amplitude across the gas diode during runaway electron-beam generation.

Using a copper cathode, which has a higher atomic number than a stainless steel cathode, did not increase the beam current amplitude (see Fig.~5 and Tables~2 and 3 in \cite{Zhang_2013}) because it reduced the voltage amplitude across the gap. The SLEP-150M generator voltage was the same in these experiments. The authors of \cite{Pasko_2024} also did not consider the second result presented in Fig.~8 of \cite{Zhang_2013}. When the voltage pulse rise time was increased to 15~ns with a 50~ns FWHM, the breakdown voltage and current in the gas-filled diode no longer depended on the cathode material (Fig.~7 in \cite{Zhang_2013}). Under these conditions, the X-ray exposure dose also did not depend on the atomic number of the cathode material.

\section{\label{sec:level5}Conclusions}
The work has shown that in accelerators with a vacuum diode, when using cathodes with a large effective area (hundreds of cm\textsuperscript{2}) and voltage pulses of long duration (units of microseconds or more), which are the most suitable for realizing the mechanism of photoelectric absorption of bremsstrahlung \cite{Pasko_2023, Pasko_2024}, the percentage fraction of AEE at voltages of hundreds of keV amounts to fractions of a percent. On an accelerator with an effective cathode area of 500~cm\textsuperscript{2} at a voltage across the vacuum diode of 160~kV, the AEE fraction did not exceed 0.7~\% and decreased severalfold when the voltage was lowered to 
110~kV.

It has been established that the largest contribution of electrons with anomalous energy to the beam current of accelerators with vacuum diodes operating at hundreds of kilovolts is achieved with short pulse durations and rise times. With a 2~ns FWHM and a 0.5~ns rise time, the AEE fraction was 24~\%. As shown in \cite{Boichenko_2011, Kozhevnikov_2022} and confirmed in the present work, AEEs in accelerators with vacuum diodes are generated at the front edge of the voltage pulse, during a rapid redistribution of the electric field in the diode due to the high space-charge density of the emitted electrons.

\begin{acknowledgments}
This research was supported by the State assignment program of the ISE~SB~RAS, project No.~FWRM-2026-0008, FWRM-2026-0009.

The authors also wish to express their sincere appreciation to Prof.~Dr.~Vladislav~Igumnov (School of Physics, Harbin Institute of Technology), whose provision of computational resources and ongoing informational support were instrumental in carrying out this research despite all the difficulties encountered.
\end{acknowledgments}

\bibliography{AEE}

\end{document}